\documentclass[aps,pra,twocolumn,superscriptaddress]{revtex4-1}
\usepackage{graphicx}
\usepackage{hyperref}
\usepackage{amsmath}
\usepackage{amssymb}
\usepackage{epstopdf}
\usepackage{multirow}
\usepackage{threeparttable}
\usepackage[usenames]{color}

\begin{document}

\title{Mid-infrared single-photon sub-pixel temporal ghost imaging}
\author{Wen Zhang}
\affiliation{State Key Laboratory of Precision Spectroscopy, and Hainan Institute, East China Normal University, Shanghai 200062, China}
\author{Kun Huang}
\email{khuang@lps.ecnu.edu.cn}
\affiliation{State Key Laboratory of Precision Spectroscopy, and Hainan Institute, East China Normal University, Shanghai 200062, China}
\affiliation{Chongqing Key Laboratory of Precision Optics, Chongqing Institute of East China Normal University, Chongqing 401121, China}
\affiliation{Collaborative Innovation Center of Extreme Optics, Shanxi University, Taiyuan, Shanxi 030006, China}
\author{Zhibin Zhao}
\affiliation{State Key Laboratory of Precision Spectroscopy, and Hainan Institute, East China Normal University, Shanghai 200062, China}
\author{Huijie Ma}
\affiliation{State Key Laboratory of Precision Spectroscopy, and Hainan Institute, East China Normal University, Shanghai 200062, China}
\author{Ziyu He}
\affiliation{State Key Laboratory of Precision Spectroscopy, and Hainan Institute, East China Normal University, Shanghai 200062, China}
\author{Jianan Fang}
\affiliation{State Key Laboratory of Precision Spectroscopy, and Hainan Institute, East China Normal University, Shanghai 200062, China}
\affiliation{Chongqing Key Laboratory of Precision Optics, Chongqing Institute of East China Normal University, Chongqing 401121, China}
\author{Heping Zeng}
\email{hpzeng@phy.ecnu.edu.cn}
\affiliation{State Key Laboratory of Precision Spectroscopy, and Hainan Institute, East China Normal University, Shanghai 200062, China}
\affiliation{Chongqing Key Laboratory of Precision Optics, Chongqing Institute of East China Normal University, Chongqing 401121, China}
\affiliation{Shanghai Research Center for Quantum Sciences, Shanghai 201315, China}
\affiliation{Chongqing Institute for Brain and Intelligence, Guangyang Bay Laboratory, Chongqing, 400064, China}

\begin{abstract}
	
Temporal ghost imaging (TGI) enables ultrafast temporal signal recovery using slow detectors, offering a promising route for high-speed mid-infrared (MIR) detection. However, conventional schemes remain limited in temporal resolution by the modulation bandwidth or pattern timescale, and are mostly confined to structured illumination. Here, we demonstrated a high-resolution MIR single-photon computational TGI system, which integrated nonlinear structured detection with sub-pixel temporal shifting. A pre-programmed near-infrared pump serves as a temporally optical gate to drive sum-frequency generation in a nonlinear crystal. Consequently,  MIR waveforms at 3.4 $\mu$m were upconverted, and captured by a room-temperature silicon detector. We realized sub-pixel operation by fractional-bin temporal stepping of the gate and multi-shot fusion via pseudo-inverse reconstruction. The sub-pixel shifting strategy decouples the achievable resolution from modulation speed, enabling 40 ps temporal precision at a driving rate of only 3.125 Gbps. This performance surpasses both detector jitter and pattern-rate limits, while maintaining single-photon sensitivity. The presented paradigm establishes a versatile route for ultrafast MIR waveform reconstruction, opening new opportunities in high-resolution infrared sensing and quantum photonics.

\end{abstract}

\maketitle

\section{Introduction}

The mid-infrared (MIR) band features molecular fingerprint signatures, atmospheric transparency, and scattering resilience \cite{Vodopyanov2020Book}. These properties make it indispensable for ultrafast molecular spectroscopy \cite{Cheng2015Science, Qian2025NM}, thermal imaging and target recognition \cite{Huang2024NC}, as well as free-space optical communication \cite{Fang2020NC, Zou2022NC}. Nowadays, the growing demand for real-time sensing and data transfer imposes stringent requirements on the operation bandwidth and detection sensitivity of MIR detectors. However, conventional MIR detectors based on narrow-bandgap semiconductors such as HgCdTe and InSb are hindered by long carrier lifetimes, dark current, and thermal noise, which limit their bandwidth to below the GHz level and degrade room-temperature sensitivity \cite{Di2025AM}. Advanced devices including quantum cascade detectors (QCDs) and quantum well infrared photodetectors (QWIPs) can reach multi-GHz or tens-of-GHz bandwidth \cite{Yang2025IPT,  Palaferri2018Nature}, while superconducting nanowire single-photon detectors (SNSPDs) achieve single-photon detection \cite{Chen2021SB, Taylor2023Optica, Pan2022OE}, yet these approaches face practical barriers such as low absorption efficiency, cryogenic operation, or restricted wavelength coverage. Despite progress with emerging two-dimensional materials \cite{Nowakowski2025science}, room-temperature MIR detection with both high sensitivity and high temporal resolution remains an unsolved challenge.

Alternatively, temporal ghost imaging (TGI) has recently emerged as an indirect detection technique that relaxes the bandwidth demands of ultrafast measurements \cite{Ryczkowski2016NP}. Instead of using high-speed detectors, TGI reconstructs temporal profiles by correlating the integrated intensity of the `bucket' detector with reference fluctuations or pre-encoded patterns. Various extensions, including differential \cite{Oka2017APL}, Fourier \cite{Meng2020OLE}, and magnified TGI \cite{Ryczkowski2017APLP}, have been introduced to improve resolution, enrich spectral information, or boost signal-to-noise ratio. Inspired by spatial-domain ghost imaging \cite{Erkmen2010AOP, Li2024COL, Salem2013AOP}, computational TGI \cite{Lantz2016Optica, Xu2018OE} has been implemented to combine a single slow detector with programmable patterns. This approach shows strong noise immunity \cite{Ferri2005PRL} and has found applications in secure communication \cite{Wang2020OLE, He2025APLP}, underwater links \cite{Chen2021OL}, and ultrafast laser diagnostics \cite{Ratner2019PRX}.

Despite its potential, most TGI studies remain in the visible and near-infrared (NIR) bands. Extending TGI to the MIR is highly desirable \cite{Wu2024LSA} but faces challenges from immature modulator and detector technologies. Frequency conversion has been employed to bypass slow MIR detectors by upconverting signals into faster near-infrared bands \cite{Wu2019Optica}. Pre-programmed NIR patterns can also be transferred to the MIR via difference-frequency generation \cite{Wu2024LSA}, while frequency upconversion has been exploited to reach single-photon sensitivity \cite{Zhang2025LPR}. Yet prior schemes mostly rely on structured illumination, which becomes impractical when the MIR source cannot be modulated or originates from natural emissions. This highlights the need for time-domain structured detection, analogous to single-pixel imaging in the spatial domain \cite{Edgar2019NP, Song2025LPR}. Moreover, the resolution of TGI is fundamentally limited either by the modulation speed in pre-programmed schemes \cite{Wu2024LSA, Zhang2025LPR} or by the characteristic fluctuation scale in random-source schemes \cite{Ryczkowski2016NP, Ryczkowski2017APLP, Wu2019Optica}. Although modern electro-optic modulators (EOMs) achieve rates $>$100 Gbps \cite{Mao2024CM}, they require costly high-bandwidth RF sources. Digital micromirror devices offer femtosecond-level control \cite{Zhao2021Optica}, but are constrained by pixel count. These trade-offs underscore the urgency of developing new structured detection strategies that can deliver high temporal resolution without relying solely on extreme modulator speeds.

In this work, we extend sub-pixel ghost imaging concepts \cite{Phillips2017SA, Tetsuno2017OE} into the temporal domain and demonstrate MIR single-photon computational TGI with sub-pixel resolution. Using a NIR EOM to generate programmable gating pattern, we implement nonlinear structured detection via optical parametric interaction in a nonlinear crystal. The MIR temporal waveforms at 3.4 $\mu$m are upconverted to 0.8 $\mu$m and captured by a room-temperature silicon detector. By applying sub-pixel temporal stepping to the pump pattern, the temporal resolution is decoupled from modulation speed, which enables 40 ps resolution at a driving rate of only 3.125 Gbps. This overcomes both the timing-jitter limit of single-photon detectors and the intrinsic resolution limit of pre-programmed patterns, establishing a robust framework for high-resolution MIR waveform reconstruction at the single-photon level.

\begin{figure*}[t!]
	\includegraphics[width=0.85\textwidth]{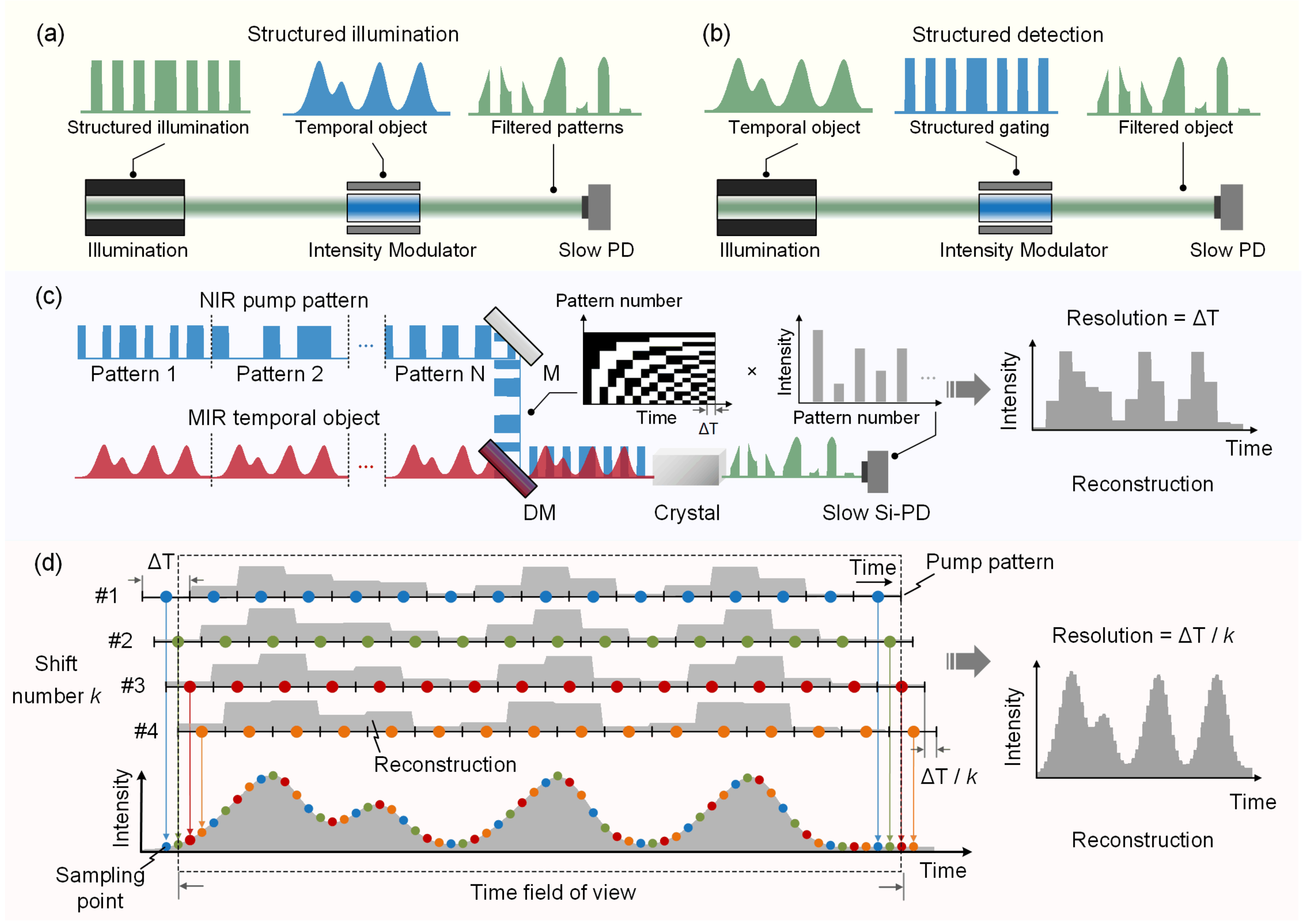}
	\caption{Conceptual illustration of MIR sub-pixel temporal ghost imaging. (a) Structured illumination configuration of the TGI system. (b) Structured detection configuration of the TGI system. (c) Nonlinear structured detection scheme based on sum-frequency generation, where pre-programmed near-infrared pattern act as a temporal gate to sample the MIR signal in a nonlinear crystal. The upconverted light is collected by a slow silicon photodetector, and the temporal object is reconstructed via inverse algorithms. (d) Principle of sub-pixel shifting operation. The pump pattern is incrementally shifted relative to the fixed temporal object along the time axis, which enables higher effective temporal resolution in the reconstruction.}
	\label{fig1}
\end{figure*}

\section{Basic principle}
Figure \ref{fig1} illustrates the concept of the proposed MIR sub-pixel TGI system, which is built on two key components: nonlinear structured detection and sub-pixel displacement. Similar to single-pixel imaging in the spatial domain, TGI can be configured in two ways: structured illumination and structured detection \cite{Edgar2019NP}, as shown in Figs. \ref{fig1}(a-b). In the structured illumination mode, a pre-programmed pattern is imposed on the light source, and the integrated intensity of the modulated probe is measured for reconstruction. In structured detection, the pre-programmed pattern is applied as a temporal gate after the object, and the gated signal is recorded directly. While the optical layouts differ, the two strategies share the same underlying principle.

For MIR applications, structured illumination can be realized with chaotic light sources exhibiting random intensity fluctuations \cite{Wu2019Optica} or by using difference-frequency generation (DFG) \cite{Wu2019Optica, Zhang2025LPR}. However, this approach becomes infeasible when the MIR source cannot be modulated or when its intrinsic temporal structure must be studied. In such cases, structured detection offers a more practical solution \cite{Wang2023NC}. To this end, we developed a nonlinear structured detection scheme based on sum-frequency generation (SFG), as shown in Fig. \ref{fig1}(c). A pre-programmed temporal pattern is encoded onto a NIR pump beam and combined with the MIR signal via a dichroic mirror. The two beams are co-focused into a periodically poled lithium niobate (PPLN) crystal to perform the SFG process. Under proper phase-matching conditions, the upconverted intensity is given by
\begin{equation}
	I_{\mathrm{up}}(t) \propto I_s(t) \cdot I_p(t) \ ,
	\label{eq1}
\end{equation}
where $I_s(t)$ and $I_p(t)$ denote the temporal intensities of the MIR signal and the NIR pump, respectively. The pump pattern therefore functions as a structured temporal gate, and the upconverted light is collected by a silicon photodetector to record the gated intensity of the temporal object.

Assuming that the temporal object is represented by $\mathcal{O} \in \mathbb{R}^{N \times 1}$, the measurement can be expressed as $B = \Phi \mathcal{O}$, where $B \in \mathbb{R}^{N \times 1}$  represents the integrated intensity measured by the bucket detector, and $\Phi \in \mathbb{R}^{N \times N}$ is a Hadamard matrix enabling efficient reconstruction \cite{Zhao2021Optica}. To mitigate source fluctuations, complementary pattern pairs are used with differential acquisition \cite{Edgar2019NP, Zhao2021Optica}. The temporal waveform is then retrieved via matrix inversion:
\begin{equation}
{O}(t) = \Phi^{-1} B \ ,
\label{eq2}
\end{equation}
where $O \in \mathbb{R}^{N \times 1}$ is the reconstructed temporal object. Here the temporal resolution is determined by the time pixel size $\Delta T$ of the measurement matrix. Traditionally, improving resolution requires higher modulation rates, which demands expensive and technically challenging high-speed devices.

To overcome this limitation, we extended the concept of sub-pixel ghost imaging from the spatial to the temporal domain \cite{Phillips2017SA, Tetsuno2017OE}, as shown in Fig. \ref{fig1}(d). Each temporal pixel corresponds to a sampling bin of width $\Delta T$. Instead of increasing the modulation rate, the pump pattern is shifted relative to the temporal object in increments of $\Delta T/k$, with $k$ successive measurements. This reduces the effective sampling interval to $\Delta T/k$, and the overlap of multiple measurements encodes sub-pixel information within the temporal field of view $\mathcal{T} = N\Delta T - \Delta T + \Delta T/k$, yielding $\mathcal{N} = kN - k + 1$ sub-pixels.
 
In this case, the measurement matrix becomes a rectangular matrix $\tilde{\Phi} \in \mathbb{R}^{(kN) \times \mathcal{N}}$, and the detection process is described by $\tilde{B} = \tilde{\Phi} \tilde{\mathcal{O}}$, where $\tilde{B} \in \mathbb{R}^{(kN) \times 1}$ and $\tilde{\mathcal{O}} \in \mathbb{R}^{\mathcal{N} \times 1}$ denote the integrated intensity of the bucket detector and  temporal object, respectively. The waveform is reconstructed using a least-squares pseudoinverse:
\begin{equation}
\tilde{O}(t) = \tilde{\Phi}^{+} \tilde{B} ,
\label{eq3}
\end{equation}
where $\tilde{\Phi}^{+}$ is the Moore-Penrose pseudoinverse of $\tilde{\Phi}$, and $\tilde{O} \in \mathbb{R}^{\mathcal{N} \times 1}$ is the reconstructed temporal object with a resolution of sub-pixel size $\Delta T/k$. This method improves temporal resolution by a factor of $k$, making it dependent on the precision of temporal stepping rather than the intrinsic modulation speed.  In analogy to super-resolution imaging in the spatial domain \cite{Park2003SPM, Li2020OE}, this sub-pixel temporal operation effectively bypasses the intrinsic limits of modulation speed, providing a powerful route to enhance temporal resolution beyond conventional optical and electronic constraints \cite{Mano2019NC}.

\begin{figure}[t!]
	\includegraphics[width=1\columnwidth]{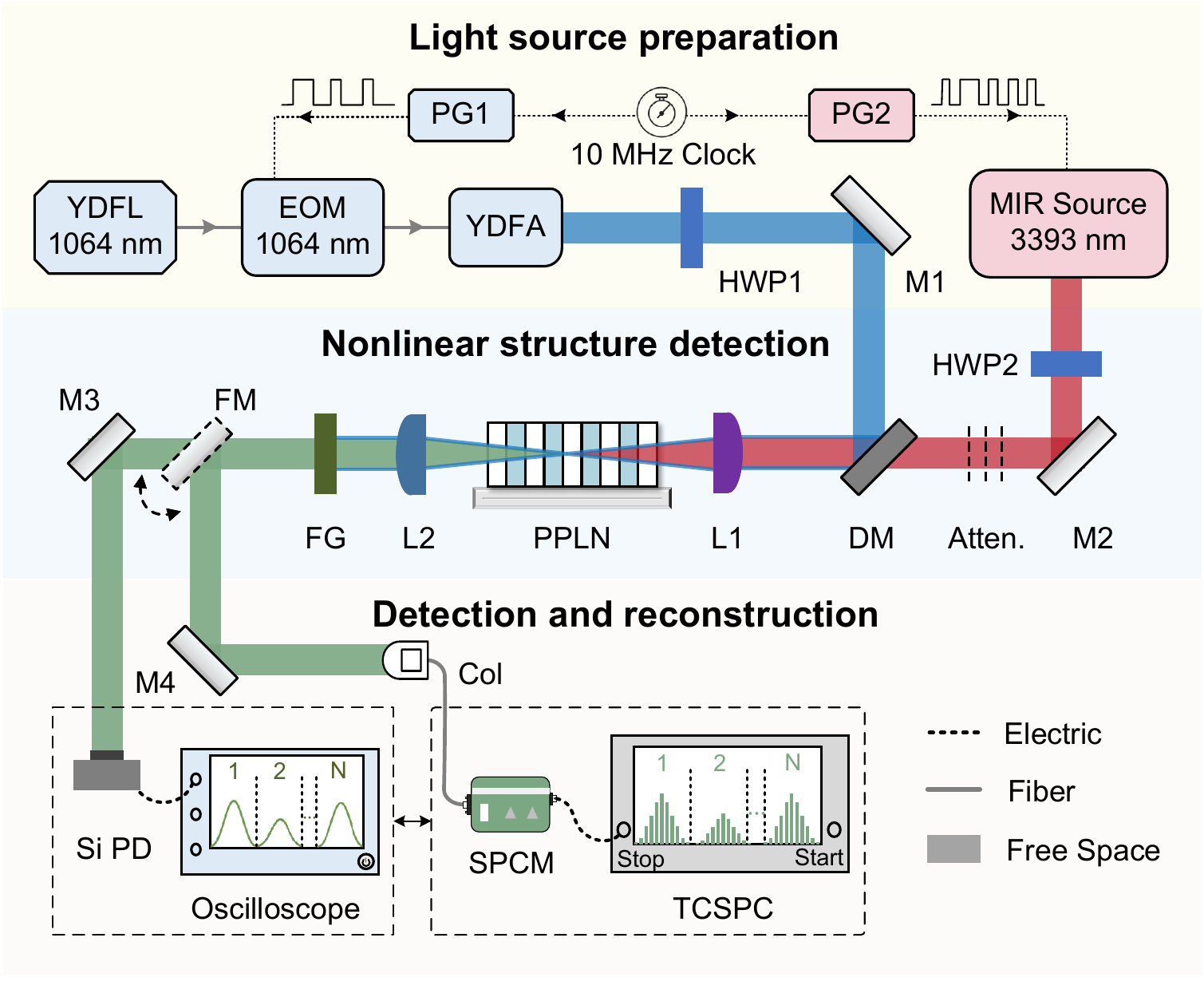}
	\caption{Experimental setup of the sub-pixel MIR TGI system. A Yb-doped fiber laser (YDFL) at 1064 nm provided the near-infrared pump, which was intensity-modulated by an electro-optic modulator (EOM) and amplified by a YDFA. The MIR source was attenuated by a neutral density filter (Atten.) and combined with the pump in a periodically poled lithium niobate (PPLN) crystal to generate upconverted light via sum-frequency generation. The upconverted signal was detected either by a silicon photodiode (Si PD) and an oscilloscope for intensity measurement, or by a single-photon counting module (SPCM) with a time-correlated single-photon counting (TCSPC) unit for photon-resolved detection. Correlation with pre-programmed temporal patterns enabled high-resolution reconstruction of the MIR waveform. PG, pattern generator; M, mirror; DM, dichroic mirror; HWP, half-wave plate; L, lens; FG, filtering group; FM, flipping mirror; Col, fiber collimator.}
	\label{fig2}
\end{figure}

\begin{figure*}[t!]
	\includegraphics[width=0.80\textwidth]{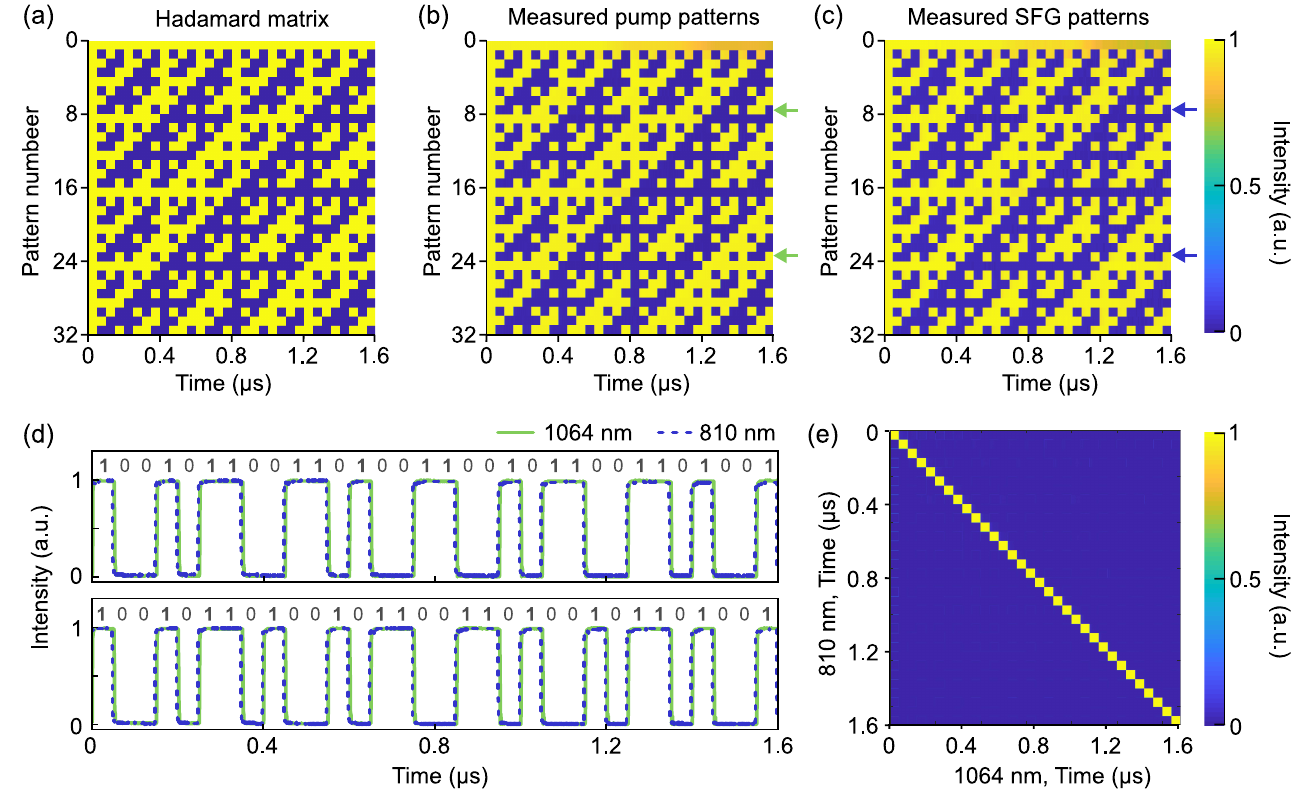}
	\caption{High-fidelity temporal mapping in the nonlinear structured detection scheme. (a) Pre-programmed Hadamard matrix loaded onto a NIR EOM at 20 Mbps. (b-c) Measured patterns of the modulated NIR light and the SFG light. (d) Two representative temporal traces of the 1-$\mu$m modulated light and the 0.8-$\mu$m upconverted light, recorded with a 0.5 GHz InGaAs photodetector. The 8th and 24th sequences of the 32-order Hadamard matrix were selected, with their binary encodings shown above. (e) Time-to-time intensity cross-correlation between the dual-color patterns, calculated over 32 distinct sequences.}
	\label{fig3}
\end{figure*}

\begin{figure*}[t!]
	\includegraphics[width=0.73\textwidth]{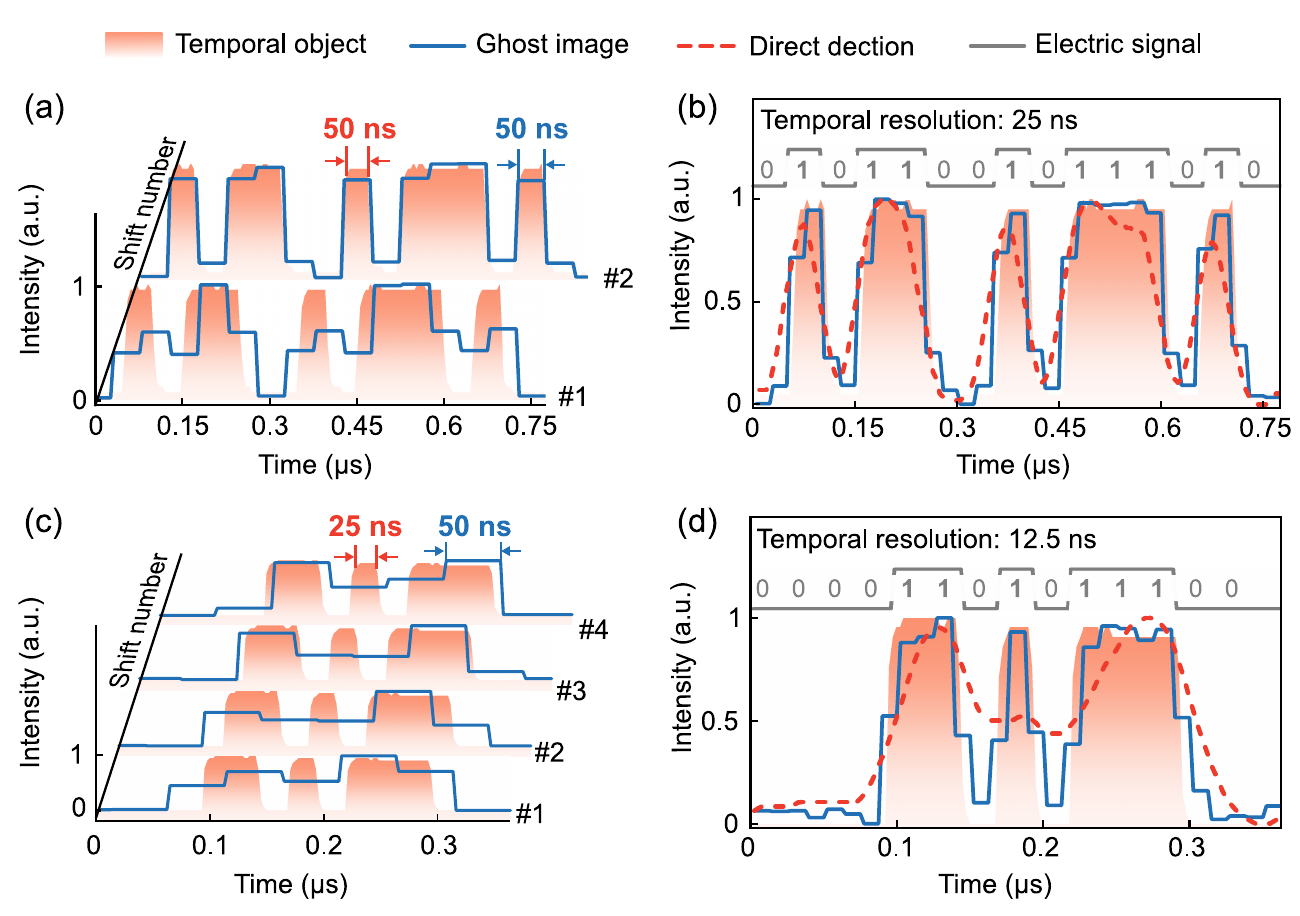}
	\caption{MIR sub-pixel TGI for binary temporal object reconstruction. (a) Ghost image (solid blue) of a 20 Mbps object in structured detection mode, compared with the ground truth (shaded red) from a 0.7 GHz MIR detector. Two measurements were taken with a 25 ns step. (b) Sub-pixel reconstruction with two shifts: the gray line shows the NIR EOM drive, and the red dashed line shows direct measurement by a 10 MHz silicon detector. (c) Reconstruction of a 40 Mbps object using 20 Mbps probe patterns with four measurements at a 12.5 ns step. (d) Sub-pixel TGI with four shifts, demonstrating an improved temporal resolution beyond the probe modulation rate.}
	\label{fig4}
\end{figure*}

\begin{figure*}[t!]
	\includegraphics[width=0.85\textwidth]{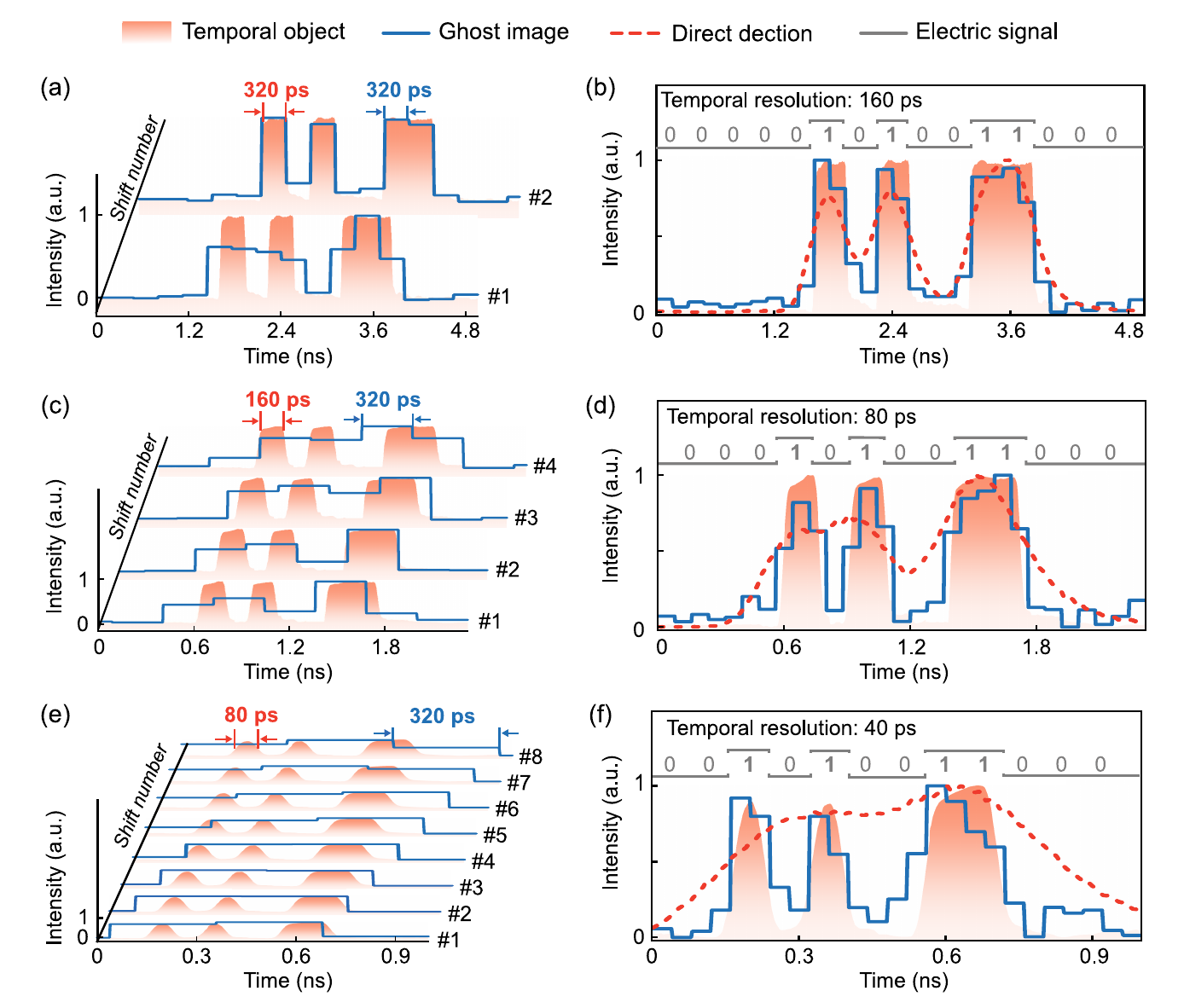}
	\caption{High-resolution MIR single-photon sub-pixel TGI. (a) TGI of a 3.125 Gbps temporal object (blue line) compared with the ground truth (red shaded). Two encoding sets were used at a 160 ps step. (b) Imaging with two sub-pixel shifts, where the gray line shows the encoded bit sequence and the red dashed line shows photon count histograms from conventional TCSPC. (c-d) Sub-pixel TGI of a 6.25 Gbps temporal object using four encoding sets with an 80 ps step. (e-f) Sub-pixel TGI of a 12.5 Gbps temporal object using eight encoding sets with a 40 ps step. An effective temporal resolution of 40 ps is achieved, surpassing the 717 ps jitter limit of the single-photon detector.}
	\label{fig5}
\end{figure*}

\section{Results and discussion}
\subsection{Experimental setup}
Figure \ref{fig2} shows the experimental setup of the MIR single-photon sub-pixel temporal ghost imaging system, which relies on nonlinear frequency upconversion for both structured detection and high-sensitivity MIR measurements. A 1064-nm Yb-doped fiber laser (YDFL) provided continuous-wave (CW) linearly polarized light at 24 mW. The light was intensity-modulated by an EOM driven by a pulse pattern generator (PG1, Keysight 81133A). The modulated pump was amplified by a Yb-doped fiber amplifier (YDFA) to 2.5 W to enhance SFG efficiency. To emulate the temporal object, a 3.4 $\mu$m MIR signal with a user-defined profile was generated via DFG between a modulated 1550 nm signal and a 1064 nm CW pump in a PPLN crystal. The 1550 nm intensity modulator was driven by PG2 (Anritsu MP1763C). Both PG1 and PG2 were synchronized to a rubidium clock to ensure long-term stability.

The NIR and MIR beams were converted to vertical polarization with half-wave plates (HWPs) to satisfy the SFG process, spatially overlapped with a dichroic mirror (DM), and focused into a 50 mm-long PPLN crystal (22.4 $\mu$m period) using a CaF$_2$ lens (LBTEK, BCX70613). The crystal temperature was stabilized at 102.6 $\pm$ 0.1 $^\circ$C. Inside the crystal, the pre-programmed NIR pump pattern served as a temporal gate for the MIR signal, and their product was mapped into upconverted light via SFG. At a pump power of 2.5 W, the nonlinear conversion efficiency was measured to be approximately 1.2\%. Further efficiency optimization can be achieved by increasing the pump power or utilizing PPLN waveguides. To suppress background noise, the signal passed through a filter assembly consisting of a 1064 nm notch filter (44-nm bandwidth), a 700 nm long-pass, a 900 nm short-pass, and an 810 nm band-pass filter (3-nm bandwidth). The filtering assembly introduced a total insertion loss of $\sim$3 dB and achieved a pump rejection ratio of $\sim$190 dB, thereby ensuring the high extinction conditions required for single-photon detection. The filtered light was then coupled into a single-mode fiber for detection.

For detection, a silicon photodiode with a 10-MHz bandwidth and a real-time oscilloscope (LeCroy Waverunner 9404M-MS) were used for intensity measurements. Under low-light conditions, a single-photon counting module (SPCM, Excelitas SPCM-AQRH-54-FC) with a time-correlated single-photon counting (TCSPC) system (Qutools, quTAG) recorded photon histograms. The temporal object was recovered by correlating the measured intensities with the pre-programmed patterns via least-squares pseudoinverse \cite{Tetsuno2017OE}.

The theoretical Hadamard matrix shown in Fig. \ref{fig3}(a) was encoded onto the NIR EOM at a modulation rate of 20 Mbps. The modulated NIR pump was detected using a 0.5-GHz InGaAs photodetector, and its time-resolved intensity [Fig. \ref{fig3}(b)] shows excellent agreement with the theoretical Hadamard sequence. This pump was then employed as a nonlinear temporal gate for structured detection. To confirm faithful information transfer from the NIR pump to the upconverted band, the MIR temporal object was blocked, and the MIR beam alone was injected into the crystal. The generated SFG light, measured by a 0.5-GHz high-speed detector [Fig. \ref{fig3}(c)], preserves the encoded temporal structure. Two representative temporal traces are shown in Fig. \ref{fig3}(d), where the dual-color waveforms overlap well in the time domain. Furthermore, the time-to-time intensity cross-correlation of 32 distinct Hadamard sequences [Fig. \ref{fig3}(e)] highlights the high-fidelity temporal mapping provided by nonlinear structured gating. Together, these results establish a robust foundation for sub-pixel TGI based on nonlinear structured detection.

\begin{table*}[t]
	\renewcommand\arraystretch{2}
	\setlength{\tabcolsep}{6pt}
	\caption{Performance comparison of representative TGI systems.}
	\label{tab1}
	\begin{tabular*}{0.85\linewidth}{@{}cccccc@{}}
		\hline
		Ref. & Wavelength  & Scheme & Modulator & Detector &  Resolution \\ 
		\hline
		This & 3.4 $\mu$m  & Structured detection & NIR EOM & NIR SPCM & 40 ps  \\
		\cite{Zhang2025LPR} & 3.4 $\mu$m  & Structured illumination & NIR EOM & NIR SPCM & 80 ps \\
		\cite{Wu2024LSA} & 3.4 $\mu$m & Structured illumination & NIR AOM & MIR  detector & 100 ns \\
		\cite{Wu2019Optica} & 2 $\mu$m & Structured illumination & / & NIR and MIR detectors &  1 ns \\
		\cite{Ryczkowski2016NP} & 1.55 $\mu$m & Structured illumination &/& NIR detectors  & 55 ps\\
		\cite{Ryczkowski2017APLP} & 1.55 $\mu$m& Structured illumination & / & NIR detectors  & 360 ps\\ 
		\hline
	\end{tabular*}
\end{table*}

\subsection{Resolution enhancement via sub-pixel operation}
Figure \ref{fig4} demonstrates the performance of MIR sub-pixel TGI for binary temporal object detection under high illumination. In Fig. \ref{fig4}(a), the NIR pump patterns were modulated at 20 Mbps and used to nonlinearly sample a 20 Mbps MIR temporal object in the PPLN crystal. The ground truth (red shaded) was obtained directly with a 0.7 GHz HgCdTe detector, while the reconstructed result (blue solid) was retrieved via matrix inversion. The temporal resolution is limited to 50 ns by the 20 Mbps modulation rate. As predicted by the Nyquist theorem, insufficient sampling leads to aliasing, and the 20 Mbps temporal structure cannot be faithfully recovered.

To improve resolution, the pump pattern was shifted by half a pixel along the time axis, and two measurements were performed. This effectively doubled the sampling points. After pseudo-inverse reconstruction, the binary encoding of the object was accurately retrieved with a resolution of 25 ns [Fig. \ref{fig4}(b)]. The gray line indicates the original electrical signal, while the red dashed curve shows direct measurement with a slow 10 MHz silicon detector for comparison.

For a higher-speed object at 40 Mbps, a 20 Mbps probe rate again results in severe under-sampling and aliasing [Fig. \ref{fig4}(c)]. By applying quarter-pixel shifts and performing four measurements, we increased the effective sampling density. The reconstructed waveform in Fig. \ref{fig4}(d) achieved a 12.5 ns resolution. These results confirm that sub-pixel TGI not only bypasses the bandwidth limitations of MIR detectors but also relaxes the stringent requirements on modulation speed, providing a practical path toward high-resolution MIR temporal imaging.

We note that sub-pixel temporal operations decouple the attainable resolution from the pattern modulation rate, shifting the primary constraint to the precision of delay stepping. While this approach necessitates multiple gate acquisitions, the associated temporal cost can be mitigated through spatial multiplexing \cite{Wan2022LSA}. By assigning respective gate positions to parallel channels, this technique transforms sequential measurements into a parallelized process, substantially accelerating acquisition.

\subsection{High-resolution MIR single-photon TGI}
Finally, we investigated the performance of MIR sub-pixel TGI under single-photon-level illumination. At extremely low flux, the incident MIR signal suffers from severe photon starvation. By combining coarse attenuation using neutral density filters with fine-tuning of the pump power in the DFG unit \cite{Liu2024APN}, the MIR signal intensity was set to 0.02 photons/bit, with a corresponding total incident photon flux of $6.25 \times 10^6$ photons/pattern. For single-photon detection, the original detector was replaced by a Geiger-mode SPCM with 60\% detection efficiency at 810 nm and 717 ps time jitter. Each photon was converted into an electrical pulse, and a TCSPC system recorded photon arrival events within a defined integration window.

Conventional temporal analysis of single-photon waveforms relies on TCSPC histograms, which provide picosecond-level time calibration but are fundamentally limited by detector jitter, restricting the effective resolution to the sub-nanosecond regime \cite{Crockett2022LPR}. Previous studies attempted to circumvent this limitation by employing logic gate encoding in the NIR band, which combines simulated MIR temporal objects with probe sequences, thereby enabling TGI beyond the limitations of detector jitter \cite{Zhang2025LPR}. However, experimental validation on real high-speed MIR temporal objects has remained elusive. Here, the PG2 was used to drive a high-speed NIR EOM, generating MIR temporal object at rates up to 12.5 Gbps. Exploiting the built-in 1-ps delay resolution of the PG1, we implemented sub-pixel temporal ghost imaging. Although the pump modulation rate was limited to 3.125 Gbps by the operation bandwidth of PG1, the proposed sub-pixel approach allowed us to measure high-speed MIR temporal objects with an effective resolution of 40 ps without requiring additional hardware.

Figure \ref{fig5} summarizes the experimental results. For a 3.125 Gbps MIR temporal object probed with a 3.125 Gbps pump pattern, two measurements with a 160 ps step size doubled the resolution, as shown in Figs. \ref{fig5}(a-b). The dashed curve in Fig. \ref{fig5}(b) shows the profile retrieved by direct TCSPC detection, which is clearly inferior. As the object rate increased to 6.25 Gbps and 12.5 Gbps, TCSPC histograms became dominated by detector jitter and failed to resolve the binary structure. In contrast, sub-pixel TGI reconstructed the waveforms with 40 ps resolution [Fig. \ref{fig5}(f)], surpassing both the 717-ps jitter limit of the SPCM and the 320-ps modulation speed of the pump pattern. These results confirm that sub-pixel TGI effectively mitigates the dual bottlenecks of detector jitter and modulator bandwidth, enabling high-resolution MIR temporal imaging at the single-photon level.

\section{Conclusion}

Table \ref{tab1} summarizes representative advances in the TGI framework. As an emerging indirect detection technology, TGI has relieved the stringent bandwidth requirements for high-speed signal detection over the past decade. With the introduction of frequency-conversion strategies, its operation has expanded from the NIR \cite{Ryczkowski2016NP, Ryczkowski2017APLP} to the MIR regime \cite{Wu2019Optica, Wu2024LSA, Zhang2025LPR}, effectively addressing the lack of high-speed modulators and detectors and even enabling room-temperature single-photon detection \cite{Zhang2025LPR}. In terms of system design, most prior studies relied on structured illumination, whereas this work introduces a structured detection scheme. This shift greatly extends applicability, especially to scenarios where the MIR source cannot be modulated, such as in the characterization of ultrafast fluorescence dynamics \cite{Chen2017ACR}.

Despite these advances, the temporal resolution of TGI has remained constrained either by the fluctuation scale of random detection patterns \cite{Ryczkowski2016NP, Ryczkowski2017APLP, Wu2019Optica} or by the intrinsic speed of pre-programmed modulation \cite{Wu2024LSA, Zhang2025LPR}. To address this limitation, we introduce a sub-pixel shifting strategy that decouples temporal resolution from modulation rate. By applying fractional temporal displacements of the pump pattern, we achieved a resolution of 40-ps, exceeding both the bandwidth limits of MIR detectors and the jitter of single-photon detectors. This represents a significant advance in time-domain analysis, enabling high-resolution measurements under modest modulation and detection bandwidths. Further performance gains could be achieved through lower-jitter timing electronics and optimized clock synchronization, while reducing the sub-pixel step size would permit even finer temporal sampling. Nevertheless, the practical resolution is ultimately governed by system-level stability, the fidelity of the nonlinear gating process, and photon-counting noise.

In summary, we have demonstrated a MIR sub-pixel temporal ghost imaging scheme that integrates nonlinear structured detection with sub-pixel temporal operations at the single-photon level, which enables precise reconstruction of ultrafast temporal object with picosecond-scale resolution. This approach significantly alleviates the stringent requirements on the timing jitter of single-photon detectors and the operation bandwidth of modulation devices. Notably, with suitable nonlinear crystals such as AgGaS\textsubscript{2} \cite{Rodrigo2021LPR} or GaP \cite{Fandio2024OL}, the presented approach can be extended into the long-wave infrared and terahertz regimes, where ultrafast time-domain measurements are currently challenging. We anticipate that the demonstrated single-photon sub-pixel TGI paradigm will open new opportunities for various applications, such as long-distance free-space optical communications \cite{Zou2022NC}, high-sensitivity MIR fluorescence characterization \cite{Chen2017ACR}, and time-resolved photoluminescence spectroscopy \cite{Julsgaard2020PR}.

\section*{Acknowledgements}
This work was funded by Shanghai Pilot Program for Basic Research (TQ20220104); National Natural Science Foundation of China (62175064, 62235019, 62035005); Innovation Program for Quantum Science and Technology (2023ZD0301000); Shanghai Municipal Science and Technology Major Project (2019SHZDZX01); Natural Science Foundation of Chongqing (CSTB2025NSCQ-GPX0443); Postdoctoral Fellowship Program and China Postdoctoral Science Foundation (GZC20250545, 2024M760918, 2025T180224); Fundamental Research Funds for the Central Universities.

\section*{Conflict of Interest}
The authors declare no conflict of interest.

\section*{Data Availability Statement}
The data that support the findings of this study are available from the corresponding author upon reasonable request.

\section*{Keywords}
temporal ghost imaging; mid-infrared detection; single-photon detection; frequency upconversion detection; sub-pixel shift

%\newpage

\end{document}